\documentstyle[prd,aps,epsf,amsmath,amsthm,amssymb]{revtex}

\newcommand{\half}{\mbox{$\textstyle \frac{1}{2}$}}
\newcommand{\quat}{\mbox{$\textstyle \frac{1}{4}$}}

\begin{document}

\tighten
\draft
\preprint{DAMTP-1999-??}
\twocolumn[\hsize\textwidth\columnwidth\hsize\csname
@twocolumnfalse\endcsname

\title{Experimental Tests for Stochastic Reduction Models}

\author{
Dorje C. Brody$^{*}$ and Lane P. Hughston$^{\dagger}$ }

\address{* The Blackett Laboratory, Imperial College,
London SW7 2BZ, UK}

\address{$\dagger$ Department of Mathematics, King's College
London, The Strand, London WC2R 2LS, UK}

\date{\today}

\maketitle

\begin{abstract}
Stochastic models for quantum state reduction give rise to
statistical laws that are in many respects in agreement with those
of standard quantum measurement theory. Here we construct a
counterexample involving a Hamiltonian with degenerate eigenvalues
such that the statistical predictions of stochastic reduction
models differ from the predictions of quantum measurement theory.
An idealised experiment is proposed whereby the validity of these
predictions can be put to the test.
\end{abstract}


\vskip2pc]

Stochastic extensions of the Schr\"odinger equation have recently
attracted attention as plausible models for state vector reduction
in quantum mechanics. The purpose of this article is to propose an
experiment to test the predictions of such models, and in
particular to distinguish them from the consequences of the
projection postulate of quantum measurement theory.

According to standard quantum mechanics, the evolution of the
state of an isolated quantum system is described by a
deterministic unitary transformation, governed by the
Schr\"odinger equation.  The behaviour of the state of a quantum
system following the measurement of an observable is less well
understood, however, and has been the subject of many debates. In
standard quantum measurement theory it is usually assumed that
when the measurement of an observable with a discrete spectrum is
carried out on a system prepared in a prescribed initial state,
then the state reduces, in a nondeterministic manner, as a
consequence of the measurement, to one of the eigenstates of the
observable being measured. The precise mechanism according to
which the state reduction occurs, if indeed this is the correct
picture, is not specified in the standard theory. We can,
nevertheless, predict the statistical properties of the
measurement outcome, and for an appropriate energy scale the
standard theory provides accurate results for a wide range of
phenomena.

There have been attempts to extend the dynamics of standard
quantum mechanics in such a way that the reduction process itself
takes the form of a dynamical process. Notable amongst these are
the stochastic models for state reduction
\cite{grw,diosi,gisin,percival,hughston,adler2}. In such models,
the quantum dynamical law is governed by a combination of a
deterministic drift term and a nondeterministic fluctuating term,
driven by a standard Wiener process. For a recent review and a
comprehensive list of references, see for example \cite{pearle0}.
We shall consider here, in particular, stochastic models for the
collapse of the wave function to eigenstates of observables
commuting with the Hamiltonian. This family of models has been
studied extensively, and can be viewed as an important prototype
for state reduction models in general
\cite{gisin,percival,hughston,adler2,pearle0,brody}.

We shall say that an energy based stochastic reduction model is
{\sl admissible} if the following properties hold. Firstly, we
require that the dynamics should be norm preserving. Secondly, we
require that the expectation value of any observable that commutes
with the Hamiltonian should be a weakly conserved quantity, i.e.,
that it should satisfy the {\sl martingale} condition. Finally, we
require that the variance of any such observable should be a {\sl
supermartingale}, i.e., that it should, on average, be a
decreasing process, to ensure reduction to an eigenstate. It has
been demonstrated recently \cite{adler} that there are infinitely
many admissible stochastic models in this sense. Specific models
within this class have been investigated in great detail, and it
has been established for these models that, remarkably, the
probability of reaching a designated eigenstate is given by the
Dirac transition amplitude for the given initial state and final
eigenstate, in agreement with the predictions of standard quantum
measurement theory \cite{gisin,hughston,adler2}.

For the class of admissible stochastic models, we can now enquire
whether the resulting statistical predictions are entirely
consistent with quantum measurement theory. In what follows we
shall construct an example of a subtle instance for which the two
theories give rise to different predictions.

In order to illustrate the idea, we consider the example of a
system of two noninteracting distinguishable spin-$\frac{1}{2}$
particles in a constant applied magnetic field. The Hamiltonian of
the system is given, in suitable units, by
\begin{eqnarray}
{\hat H} = \left( {\hat S}_{1z}\otimes\hat{{\bf 1}}_{2}\right)
\oplus \left(\hat{{\bf 1}}_{1} \otimes {\hat S}_{2z}\right),
\label{eq:1}
\end{eqnarray}
where ${\hat S}_{iz}$ is the component of the $i$-th spin operator
in the direction of the applied field, which is assumed to be
oriented along the $z$-axis. The Hamiltonian $\hat{H}$ has three
eigenvalues: $+1$ corresponding to the
$|\!\!\uparrow\uparrow\rangle$ state, $-1$ corresponding to the
$|\!\!\downarrow\downarrow\rangle$ state, and $0$ corresponding to
a degenerate family of states expressible in the form $\alpha
|\!\!\uparrow\downarrow\rangle+\beta
|\!\!\downarrow\uparrow\rangle$ with $|\alpha|^2+|\beta|^2=1$.

Given the Hamiltonian above, we consider making a measurement of
the energy of the system. Now, an energy measurement is not quite
the same as a spin measurement, even though a measurement of
${\hat S}_{1z}$ and ${\hat S}_{2z}$ will certainly determine the
value of the energy. According to quantum measurement theory, a
generic initial state collapses to one of three possibilities,
namely, either to one of the energy eigenstates corresponding to
the energy eigenvalues $1$ or $-1$, or to a third energy
eigenstate with eigenvalue $0$. The standard projection postulate
\cite{neumann,luders} states that there exists a unique
decomposition of a given initial state in terms of possible
outcome states of the measurement. Therefore, in the present
example, the third eigenstate onto which the collapse may occur
can be determined uniquely in terms of the specified initial state
and two remaining nondegenerate eigenstates.

In order to clarify the situation, we examine the case where the
initial state is given by
\begin{eqnarray}
|\psi_{I}\rangle = \half \left( |\!\!\uparrow\uparrow\rangle +
|\!\!\downarrow\downarrow\rangle + |\!\!\uparrow\downarrow\rangle
- |\!\!\downarrow\uparrow\rangle \right) . \label{eq:2}
\end{eqnarray}
Then, according to quantum measurement theory, it follows as a
consequence of the energy measurement that $|\psi_I\rangle$
collapses to the final state $|\psi_F\rangle =
|\!\!\uparrow\uparrow\rangle$ with probability $\frac{1}{4}$ when
eigenvalue $1$ is observed, to the final state $|\psi_F\rangle =
|\!\!\downarrow\downarrow\rangle$ with probability $\frac{1}{4}$
when eigenvalue $-1$ is observed, and to the final state
$|\psi_F\rangle = \frac{1}{\sqrt{2}}
(|\!\!\uparrow\downarrow\rangle- |\!\!\downarrow\uparrow\rangle)$
with probability $\frac{1}{2}$ when eigenvalue $0$ is observed. It
should be emphasised that the projection postulate is an axiom,
and, as such, it does not directly follow from the basic dynamical
or statistical laws of quantum mechanics.

In particular, in stochastic reduction models, when there is a
degenerate eigenvalue for the energy, the outcome for the
corresponding eigenstate is incompatible with the projection
postulate, and is given rather by a range of possible pure states,
characterised by a probability distribution over the entire family
of states sharing the given energy eigenvalue.

In the example for which the initial state is (\ref{eq:2}), if the
energy eigenvalue $0$ is observed, then the statistical properties
of the resulting pure states are represented by the following
density matrix:
\begin{eqnarray}
{\hat\rho} = \int_{0}^{2\pi}\int_{0}^{\pi} \rho(\theta,\phi)
|\theta,\phi\rangle\langle \theta,\phi| {\rm d}\theta {\rm d}\phi
. \label{eq:3}
\end{eqnarray}
Here the density function $\rho(\theta,\phi)$ determines a
probability distribution over the unit sphere parameterised by the
angular coordinates $\theta,\phi$, for which the corresponding
state $|\theta,\phi\rangle$ is given by
\begin{eqnarray}
|\theta,\phi\rangle = \cos\half \theta
|\!\!\uparrow\downarrow\rangle + {\rm e}^{{\rm i}\phi}
\sin\half\theta |\!\!\downarrow\uparrow\rangle . \label{eq:4}
\end{eqnarray}
The density function $\rho(\theta,\phi)$ depends on the initial
condition $|\psi_I\rangle$ and the specific choice of the
stochastic model. In the case where $\rho(\theta,\phi)$ is given
by the $\delta$-function
\begin{eqnarray}
\rho(\theta,\phi)= \delta(\theta-\half\pi) \delta(\phi-\pi),
\label{eq:5}
\end{eqnarray}
we recover the result of the projection postulate in the standard
theory, for which the density matrix can be expressed in the form
${\hat\rho} = |\psi_F\rangle\langle\psi_F|$, where
$|\psi_F\rangle$ is the singlet state. In general, however,
$\rho(\theta,\phi)$ differs from the expression (\ref{eq:5}).

In summary, we see that, conditional on the measurement outcome of
the energy being the degenerate eigenvalue $0$, the resulting
reduced state is given either by a unique eigenstate or by a
statistical mixture of the corresponding eigenstates, depending on
which theory is employed. Thus, in the case of a degenerate energy
eigenvalue spectrum, the two theories give rise to different
predictions.

Let us consider therefore the problem of determining how one or
the other of these predictions can be ruled out by an appeal to
observation.

For our idealised experiment, we take the state $|\psi_{I}\rangle$
in (\ref{eq:2}) to be the initial state. Then, if the outcome of
the energy measurement gives eigenvalue $0$, the state collapses,
according to the standard theory, to the spin-0 singlet state. The
stochastic theory, however, predicts that the collapse is to a
random pure state lying somewhere on the subspace consisting of
all superpositions of the $|\!\!\uparrow\downarrow\rangle$ state
and the $|\!\!\downarrow\uparrow\rangle$ state. At this point, we
may consider the possibility of using a spin measurement, say, on
the first particle, to distinguish the two possibilities. Then, if
the system is in the singlet state, the outcome of a measurement
of ${\hat S}_{1z}$ is $|\!\!\uparrow\downarrow\rangle$ with
probability $\frac{1}{2}$ and $|\!\!\downarrow\uparrow\rangle$
with probability $\frac{1}{2}$. On the other hand, according to
the stochastic theory, the probability of observing the
$|\!\!\uparrow\downarrow\rangle$ state is given by
\begin{eqnarray}
p = \int_{0}^{2\pi}\int_{0}^{\pi}\rho(\theta,\phi)\cos^2\half
\theta {\rm d}\theta {\rm d}\phi, \label{eq:6}
\end{eqnarray}
which may be different from $\frac{1}{2}$. If they are different,
then we can test the two predictions by this experiment.

The simple spin measurement we have just considered, however, does
not distinguish between the two predictions, because we require
that $p=\half$. In other words, for a {\sl viable} stochastic
model we insist that it should give the same statistical
predictions as the standard projection postulate, for the
secondary measurement we consider here. This is because we would
like the stochastic model to be compatible with the predictions of
the standard theory for commuting observables.

Notwithstanding the nontrivial constraint imposed by this
condition, we can show that admissible stochastic models are, in
fact, viable in the sense discussed above. To see this, we must
use the fact that, in an admissible reduction model, any
observable commuting with the Hamiltonian also satisfies a weak
conservation law, when the system collapses to an energy
eigenstate \cite{adler2,brody}. The argument proceeds as follows.
Let us consider more explicitly the measurement of the spin of the
first particle, characterised by the observable
\begin{eqnarray}
\hat{\Sigma}_{1z} = {\hat S}_{1z}\otimes\hat{{\bf 1}}_2.
\label{eq:7}
\end{eqnarray}
For the expectation value of $\hat{\Sigma}_{1z}$ in the initial
state $|\psi_I\rangle$, we find
\begin{eqnarray}
\langle\psi_I|{\hat\Sigma}_{1z}|\psi_I\rangle = 0 . \label{eq:8}
\end{eqnarray}
Now, the weak conservation law states that this is the same as the
terminal expectation value of $\hat{\Sigma}_{1z}$, after the
energy measurement is made. Therefore, we must have
\begin{eqnarray}
\quat \langle\uparrow\uparrow\!\!|
\hat{\Sigma}_{1z}|\!\!\uparrow\uparrow\rangle &+& \quat
\langle\downarrow\downarrow\!\!|\hat{\Sigma}_{1z}
|\!\!\downarrow\downarrow\rangle
\nonumber \\
& & \hspace{-1.5cm} + \half\int_{0}^{2\pi}\int_{0}^{\pi}
\rho(\theta,\phi) \langle\theta,\phi|\hat{\Sigma}_{1z}
|\theta,\phi\rangle {\rm d}\theta {\rm d}\phi = 0 . \label{eq:9}
\end{eqnarray}
The first two terms cancel, because of the relation
\begin{eqnarray}
\langle\uparrow\uparrow\!\!|\hat{\Sigma}_{1z}
|\!\!\uparrow\uparrow\rangle = -\langle\downarrow\downarrow\!\!|
\hat{\Sigma}_{1z}|\!\!\downarrow\downarrow\rangle. \label{eq:10}
\end{eqnarray}
On the other hand, as a consequence of (\ref{eq:4}) we find that
\begin{eqnarray}
\langle \theta,\phi|\hat{\Sigma}_{1z} |\theta,\phi\rangle =
2\cos^2\half\theta-1, \label{eq:11}
\end{eqnarray}
from which it follows that $p=\half$ and hence the viability
condition follows at once.

Therefore, in the case of an admissible reduction model, we must
make a measurement of an observable that does not commute with the
Hamiltonian in order to distinguish between the predictions of the
standard theory and the stochastic theory.

In the present context, this can be achieved, for example, by
measurement of the total spin operator ${\hat S}^{2}$ for the
given two-particle system. More specifically, we prepare an
ensemble consisting of a large number of identical states, each
given initially by $|\psi_I\rangle$, and carry out energy
measurements. About half of the results will give either $1$ or
$-1$, which we discard. The ensemble corresponding to remaining
half of the particle pairs will either be in the singlet state,
according to the standard theory, or will be in a mixed state
${\hat\rho}$ characterised by (\ref{eq:3}), according to the
stochastic theory.

We now carry out the measurement of ${\hat S}^2$ on each element
of the ensemble. If each final state $|\psi_F\rangle$ is in the
spin-$0$ singlet state, then the result of the measurement gives
\begin{eqnarray}
\langle\psi_F|{\hat S}^2|\psi_F\rangle = 0 \label{eq:12}
\end{eqnarray}
for the ensemble average. On the other hand, if the statistics of
the ensemble are given by the density matrix ${\hat\rho}$
(\ref{eq:3}), then by use of (\ref{eq:4}) we find that the
measurement results in the following ensemble average for the
total spin measurement:
\begin{eqnarray}
{\rm tr}\left({\hat\rho}{\hat S}^2\right) &=&
\int_{0}^{2\pi}\int_{0}^{\pi}
\rho(\theta,\phi)\langle\theta,\phi|{\hat S}^2|\theta,\phi\rangle
{\rm d}\theta {\rm d}\phi \nonumber \\ &=& 1 +
\int_{0}^{2\pi}\int_{0}^{\pi} \rho(\theta,\phi) \sin\theta\cos\phi
{\rm d}\theta {\rm d}\phi. \label{eq:13}
\end{eqnarray}
This expression does not vanish unless (\ref{eq:5}) holds. More
precisely, because ${\hat S}^2$ is a nonnegative operator, the
expectation of which vanishes only in the singlet state, the
expression ${\rm tr}({\hat\rho}{\hat S}^2)$ in (\ref{eq:13})
vanishes if and only if ${\hat\rho}$ is the pure state density
matrix predicted by the standard measurement theory.

We note in deriving (\ref{eq:13}) that the expectation
$\langle\theta,\phi|{\hat S}^2|\theta,\phi\rangle$ appearing in
the integrand can be calculated by use of basis rotation
\begin{eqnarray}
|\theta,\phi\rangle &=& \frac{1}{\sqrt{2}}(\cos\half\theta
+\sin\half\theta {\rm e}^{{\rm i}\phi})|t\rangle \nonumber \\ & &
+ \frac{1}{\sqrt{2}} (\cos\half\theta-\sin\half\theta {\rm
e}^{{\rm i}\phi}) |s\rangle , \label{eq:14}
\end{eqnarray}
where
\begin{eqnarray}
|t\rangle=\frac{1}{\sqrt{2}}(|\!\!\uparrow\downarrow\rangle +
|\downarrow\uparrow\rangle)
\end{eqnarray}
and
\begin{eqnarray}
|s\rangle=\frac{1}{\sqrt{2}} (|\!\!\uparrow\downarrow\rangle -
|\downarrow\uparrow\rangle)
\end{eqnarray}
are the ${\hat S}_z=0$ triplet and the singlet states,
respectively, satisfying $\langle s|{\hat S}^2|s\rangle=0$ and
$\langle t|{\hat S}^2|t\rangle=2$. Therefore, measurements of the
total spin are sufficient to distinguish the possible outcomes of
the energy measurement predicted by the two theories.

In conclusion, we have shown an example of a system characterised
by a Hamiltonian with a degenerate eigenvalue, for which the
standard measurement theory and the stochastic reduction theory
give rise to different statistical predictions. We have also
outlined the details of an idealised experiment whereby the
validity of the two predictions can be checked by examination of
the statistical properties of measurement outcomes for the total
spin operator. The reason for introducing such higher order
moments, to distinguish the two predictions, is because of the
weak conservation law satisfied by the admissible models, which
ensures that the statistical predictions of the stochastic models
must agree with those of the standard quantum measurement theory
in the case of observables commuting with the Hamiltonian.

If the prediction of the conventional quantum measurement theory
based on the projection postulate is valid, then this will rule
out a wide class of existing stochastic models. On the other hand,
if the test for the stochastic theory turns out to be affirmative,
then the implications are profound: not only would the result will
provide strong support for stochastic models in general, but it
would also allow one to test experimentally which of the many
admissible models gives the correct prediction.

DCB acknowledges financial support from The Royal Society. We are
grateful to B.~K.~Meister and K.~P.~Tod for stimulating
discussions.

$*$ Electronic mail: dorje@ic.ac.uk

$\dagger$ Electronic mail: lane.hughston@kcl.ac.uk

\end{document}